\documentclass[aps,pre,twocolumn,showpacs,preprintnumbers,superscriptaddress]{revtex4}

\usepackage{graphicx}% Include figure files
\usepackage{epsf} % Include figure files, old times style !!
\usepackage{dcolumn}% Align table columns on decimal point
\usepackage{bm}% bold math
\usepackage{amssymb}
\usepackage{amsmath}
\usepackage{amsbsy}
\usepackage{amsbsy}
\usepackage{latexsym}

\begin{document}

\title{Single tensionless transition in the Laplacian roughening model}

\author{Juan Jes\'us Ruiz-Lorenzo}
%\email[]{ruiz@fis.unex.es}
\affiliation{Departamento de F\'{\i}sica, Facultad de Ciencias,
Universidad de Extremadura, 06071 Badajoz, and Instituto de Biocomputaci\'on y
F\'{\i}sica de los Sistemas Complejos, 50009  Zaragoza, Spain}
\author{Esteban Moro}
%\email[]{emoro@math.uc3m.es}
\affiliation{Departamento de Matem\'aticas and Grupo Interdisciplinar de Sistemas
Complejos (GISC), Universidad Carlos III de Madrid, Avenida de la Universidad 30,
E-28911 Legan\'es, Spain}
\author{Rodolfo Cuerno
}
%\email[]{cuerno@math.uc3m.es}
\affiliation{Departamento de Matem\'aticas and Grupo Interdisciplinar de Sistemas
Complejos (GISC), Universidad Carlos III de Madrid, Avenida de la Universidad 30,
E-28911 Legan\'es, Spain}

\date{\today}

\begin{abstract}
We report large scale Monte Carlo simulations of the equilibrium
discrete Laplacian roughening (dLr) model, originally introduced as
the simplest one accommodating the hexatic phase in two-dimensional
melting. The dLr model is also relevant to surface roughening in
Molecular Beam Epitaxy (MBE). We find a single phase transition,
possibly of the Kosterlitz-Thouless type, between a flat
low-temperature phase and a rough, tensionless, high-temperature
phase. Thus, earlier conclusions on the order of the phase
transition and on the existence of an hexatic phase are seen as due
to finite size effects, the phase diagram of the dLr model being
similar to that of a continuum analog previously formulated in the
context of surface growth by MBE.
\end{abstract}

\pacs{64.60.Cn, 68.35.Ct, 68.35.Rh}
%68.35.Ct, % Interface structure and roughness
%68.35.Rh, % Phase transitions and critical phenomena (en Solid surfaces and
%          % solid-solid interfaces)
%05.45.-a  % Nonlinear dynamics and nonlinear dynamical systems
%68.55.-a, % Thin film structure and morphology
%64.60.Cn, % Order-disorder transformations; statistical mechanics of model systems
%81.15.Aa  % Theory and models of film growth
%05.50.+q  % Lattice theory and statistics (Ising, Potts, etc.)
%64.60.-i  % General studies of phase transitions
%05.10.Ln  % Monte Carlo methods
%05.70.Jk  % Critical point phenomena
%75.10.Hk  % Classical spin models
%81.10.Aj  % Theory and models of crystal growth; physics of crystal growth,
%            crystal morphology and orientation
%64.70.-p  % Specific phase transitions

\maketitle

Two-dimensional (2D) melting has played a driving r\^ole in
Statistical Physics for more than two decades. %The theoretical and experimental
Efforts made at clarifying its nature \cite{reviews_melting} have
aided to understand systems in which topological defects are
relevant, from the equilibrium fluctuations of metallic surfaces
\cite{pimpinelli} to superfluidity and superconductivity in thin
films, and phase transitions in liquid crystals \cite{nelson_book}.
One of the most intriguing related notions is the hexatic phase,
between a solid at low temperature ($T$) and an isotropic fluid at
high $T$, transitions between phases being of the
Kosterlitz-Thouless (KT) type. Such is the
Kosterlitz-Thouless-Halperin-Nelson-Young (KTHNY) mechanism for 2D
melting \cite{nelson_book}. Although controversial for some time,
the hexatic phase has indeed been found in atomistic model systems
\cite{melting_numer} and in experiments \cite{melting_exp}.

A successful approach to systems with defect-mediated phase
transitions as the above has been the use of duality to formulate
equivalent height models. E.g., the discrete Gaussian (dG) model
[Eq.\ (\ref{gral_disc_model}) below for bending rigidity parameter
$\kappa=0$] is dual of the 2D Coulomb gas, and the roughening
transition in the former corresponds \cite{chui} to the well-known KT
phase transition of the latter, driven by the unbinding of
vortex-antivortex pairs. With a similar philosophy, the discrete
Laplacian roughening (dLr) model was introduced by Nelson
\cite{nelson} to describe 2D melting. Its Hamiltonian is
\begin{equation}
{\cal H}= \frac{1}{2} \sum_\mathbf{r} \left(\sigma [\nabla_d h(\mathbf{r})]^2
+\kappa [\nabla_d^2  h(\mathbf{r})]^2 \right) \,,
\label{gral_disc_model}
\end{equation}
where $\mathbf{r}$ denotes position on a 2D lattice of lateral
size $L$, $\nabla_d$ is discrete gradient, and $h\in \mathbb{Z}$.
The original dLr model \cite{nelson} is obtained by setting to
zero the surface tension parameter $\sigma$. Note, the dLr model
is a {\em discrete} version of the {\em linear} approximation to
Helfrich's energy functional for 2D membranes \cite{safran}, and
provides a simplified description of fluctuating {\em tensionless}
surfaces, such as biological membranes \cite{safran} or, e.g.,
such as those grown under conditions typical in Molecular Beam
Epitaxy (MBE) \cite{us_prl}.

For the dLr model, the KTHNY mechanism would imply
\cite{reviews_melting} an intermediate hexatic
phase in which the surface disorders in heights, but not in slopes
(quasi-long range orientational order). For low $T$,
the surface would be in a flat phase, dual of the isotropic fluid in
melting, while for high $T$ the surface would disorder in heights and
slopes, providing the dual of the solid phase. %in the melting problem.
In terms of the surface structure factor  $S(\mathbf{q}) =
\langle \hat{h}(\mathbf{q}) \hat{h}(-\mathbf{q}) \rangle$ \cite{nota}, the
rough high $T$ phase implies power law behavior as $S(\mathbf{q})
\sim q^{-4}$, changing to $S(\mathbf{q}) \sim q^{-2}$ in the
hexatic phase \cite{note5}, and to existence of a finite
correlation length in the flat low $T$ phase. Equivalently, for
the stationary height-difference $C(r)$ and slope-difference
$C_d(r)$ correlations \cite{nota2}, these behaviors amount to:
$(i)$ rough phase $C(r) \sim r^2 \log r$, $C_d(r) \sim \log r$;
$(ii)$ hexatic phase $C(r) \sim \log r$, $C_d(r) \sim 1$; $(iii)$
flat phase $C(r) \sim 1$, $C_d(r) \sim 1$. Results supporting this
picture were obtained on small ($L\leq 32$) square and triangular
lattices \cite{bruce_et_al}. However, conflicting evidence for $L
\leq 64$ was presented that the model had a {\em single first
order} transition, see \cite{janke} and references therein. The
discrepancy has remained unsolved, in spite of recent analytical
studies \cite{us_anal}, elucidation of the phase diagram being important
to the diverse contexts mentioned above.

Here, we provide new Monte Carlo (MC) simulations of the
dLr model on the square and triangular lattices. Our results for
sizes up to $512 \times 512$, much larger than those previously
studied \cite{bruce_et_al,janke}, allow us to see previous works
as inconclusive due to finite size effects. The model has a single
continuous transition, possibly of the KT type, between the
flat and the rough phases, there being no sign of an hexatic phase
to within our numerical resolution in $T$. Notably, this provides
an instance of a roughening transition in which the rough phase
corresponds to a free {\em tensionless} surface, rather than a
free surface with tension, as in the dG model.
Moreover, the phase diagram of the dLr model is seen to resemble
closely that of a {\em continuum model} proposed
\cite{us_prl,us_anal} for MBE growth, suggesting that both
models are in the same universality class,
much like the relationship between the dG and the
continuum sine-Gordon models \cite{nozieres_review}.

For our MC simulations we follow the same procedure as in
\cite{bruce_et_al}, fluctuations being treated by
the histogram method \cite{montecarlo}, further validated through
additional simulations on different points of the extrapolated intervals.
Thermalization has been checked by monitoring the behavior of non-local
observables like the specific heat and the structure factor at the smallest
wave-vector on our finite lattices, $S(q=2 \pi/L)$, as functions of MC time. Note that
the dLr model has a richer ground state structure than the dG model,
Hamiltonian (\ref{gral_disc_model}) with $\sigma=0$ being minimized
not only by configurations with uniform heights, but also by configurations
with uniform slopes (and by more complex morphologies, see below), see e.g.\
the surface morphology made up of patches with various constant slopes shown
in Fig.\ \ref{fig4} for high $T$. In simulations, this
requires large enough system sizes and appropriate boundary conditions
so that the full minima structure can be significatively probed. In particular,
for small sizes and periodic boundary conditions the system is
effectively constrained to fluctuating around a single energy minimum
(the morphology with zero slope), inducing apparent
hystheretic behavior associated with a first order transition \cite{janke}.
In our simulations, we have employed both
periodic and free (Neumann) boundary conditions, and we have made sure that
results provided are (qualitatively) independent of these.
\begin{figure}
\begin{center}
\includegraphics[width=1.65in,clip=]{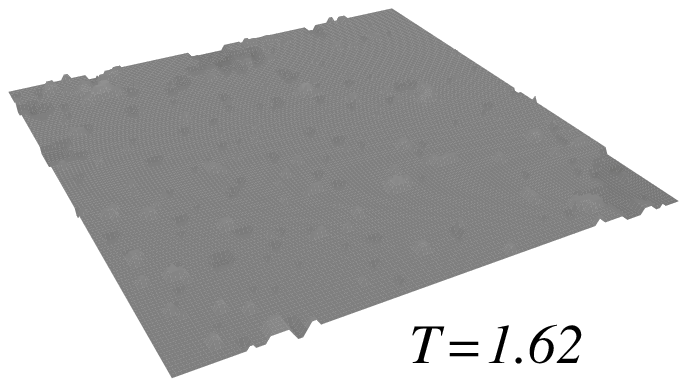}
\includegraphics[width=1.65in,clip=]{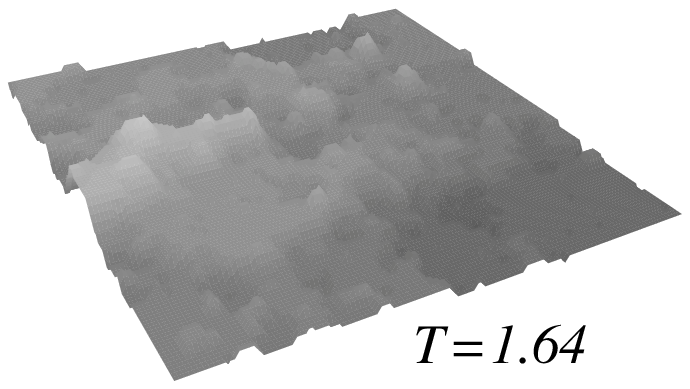}\\
\includegraphics[width=1.65in,clip=]{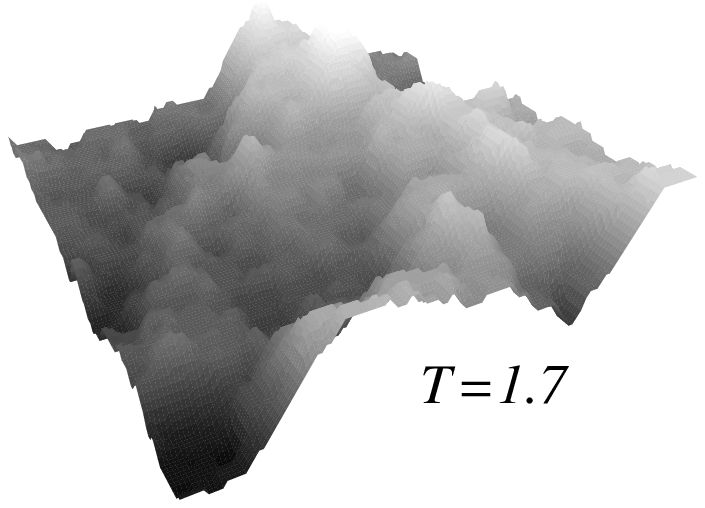}
\includegraphics[width=1.65in,clip=]{fig1d.eps}
\end{center}
\caption{Surface morphologies for three sample temperatures around
$T_c$ on the $L=128$ square lattice for Neumann (zero derivative)
boundary conditions. Inset: Lateral cut of the surface for $T=1.7$.
All units are arbitrary.} \label{fig1}
\end{figure}

As done for the dG model in \cite{nota3}, we study the phase
transition through the behavior of the structure factor $S(q)$ for
different temperatures. In order to test the KTNHY mechanism, we
have studied the behavior of $S(q)$ as a function of $T$ and $L$, by
fitting the small wave-vector part of $S(q)$ to $S(q) \sim q^{-r}$.
As seen in Fig.\ \ref{fig3} for the square lattice (for the sake of
clarity, we omit plots for the triangular lattice, in which
completely analogous results are obtained), there is no evidence of
a {\em finite} temperature interval {\em within} which $r \simeq 2$,
that would be the signature of the hexatic phase. Rather, we find a
{\em gradual} change from the flat phase behavior ($r \simeq 0$) to
the rough phase one ($r \simeq 4$). This change becomes more abrupt
when the system size is increased, so that only the flat and the
rough phases remain well-defined in the thermodynamic limit. These
results may thus explain the apparent observation of an hexatic
phase in \cite{bruce_et_al} for small $L$ values, where no
systematic finite size effects were assessed. By defining the
critical temperature $T_c$ as the value at which curves for
different system sizes cross \cite{montecarlo}, we estimate $T_c =
1.65(1)$ for the square lattice and $T_c = 1.90(2)$ for the
triangular lattice.
\begin{figure}
\begin{center}
\includegraphics[width=0.45\textwidth,clip=]{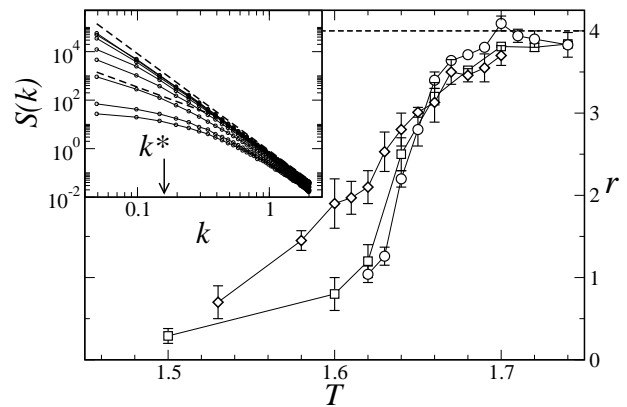}
\end{center}
\caption{Effective exponent $r$ in the small wave-vector behavior
$S(k) \sim k^{-r}$ (for $k \equiv 2 \sin(q/2) < k^*(L) \approx
3\pi/L$), as a function of $T$, for $L=32$ ($\Diamond$), 64
($\Box$), and 128 ({\sf o}). Inset: surface structure factor $S(k)$
on the $L=128$ square lattice vs $k$, for $T=1.62$ (bottom) up to
$T=1.69$ (top). Dashed reference lines have slopes $-2$ (bottom) and
$-4$ (top). All other lines are guides to the eye. All units are
arbitrary.} \label{fig2}
\end{figure}

\begin{figure}
\begin{center}
\includegraphics[width=0.45\textwidth,clip=]{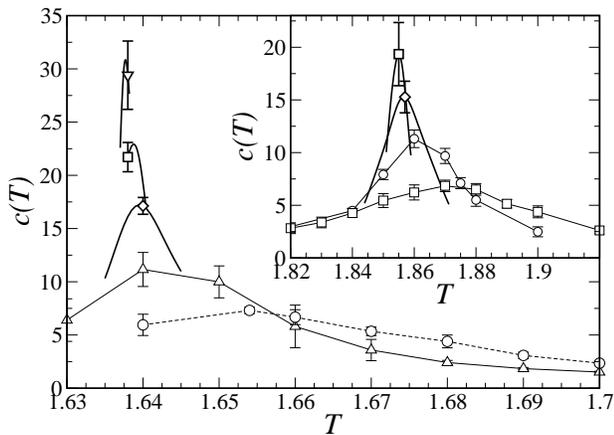}
\end{center}
\caption{$c(T)$ vs $T$ on the square (main panel) and triangular
(inset) lattices for $L=16$ ($*$), 32 ($\triangle$), 64 ({\sf o}),
128 ($\Diamond$), 256 ($\Box$), and 512 ($\bigtriangledown$). Bars
are statistical errors and thin lines are guides to the eye. For the
larger $L$ values in each case, thick solid lines show the $c(T)$
curve extrapolated by the histogram method \cite{montecarlo}. All
units are arbitrary.} \label{fig3}
\end{figure}
Further evidence on the existence of a single phase transition is provided by
the behavior the specific heat $c(T,L)=\left(\langle {\cal H}_{\rm
    dLr}^2\rangle - \langle {\cal H}_{\rm dLr} \rangle^2\right)/(T^2 L^2)$ as
a function of temperature.  Fig.\ \ref{fig1} shows $c(T)$ on the square and
triangular lattices for the largest system sizes in our simulations. Within
our statistics, a {\em single} peak at $T=T^*$ can be detected, rather than
two as would be expected within the KTHNY scenario.  The height and position
of the peak are functions of lattice size $L$. Fig.\ \ref{fig2} (inset)
provides the results of finite size analysis on the specific heat curves, in
which the maximum value $c_{\rm max}(L)$ obtained for each lattice size is
plotted as a function of $L$. Remarkably, although for lattice sizes $L
\lesssim 70$ the specific heat grows approximately as $c_{\rm max} \lesssim L$
---compatible with claims on the apparent weakly first order character of the
transition for $L\leq 64$ \cite{janke}---, for larger $L$ values the
increase of $c_{\rm max}(L)$ slows down.  For our largest simulated
systems, the best fit is logarithmic $c_{\rm max} \sim \log L$.
Actually, for the 2D XY model the specific heat at the transition
temperature is known \cite{himbergen} to first grow logarithmically
with system size and then saturate for large enough values of $L$,
suggesting our result might reflect finite size effects. Indeed,
saturation is expected provided that the correlation length at
$T^{*}$ is smaller than $L$ and thus a horizontal plateau would
occur at low $k$ for $S(k)$, namely $r = 0$ as defined in Fig.\
\ref{fig2}. The steady decrease of $r$ at $T^*$ for increasing $L$
indicates that such a condition has not being reached. Persistence
of logarithmic behavior in the $L\to \infty$ limit would rather
suggest that the phase transition is in e.g.\ the 2D Ising class
\cite{chaikin}. In order to explore this possibility, in Fig.\
\ref{fig2} we study the dependence of the specific heat jump
position $T^*(L)$ with lateral size $L$. In a continuous transition,
$T^*(L)$ scales as~\cite{chaikin} $T^*(L)-T^* \sim L^{-1/\nu} (1 + g
L^{-\omega})$, where $g$ is a numerical constant and $\omega$ is an
exponent that accounts for corrections to scaling, and is in the
range $7/4 \leq \omega \leq 2$ for the 2D Ising class \cite{sokal}.
The best multiparameter fit to such scaling form yields $\nu =
1.54(47)$, and $\nu = 0.94(16)$ on the square and the triangular
lattices, together with $\omega= 1.6(3)$, and 2.2(2.0) respectively,
to be compared with $\nu=1$ for the 2D Ising class
\cite{nota_no_corr}.  Although these results might seem compatible
with 2D Ising universality for the present transition, we believe
our numerical evidence favors more strongly a different
interpretation.  Thus, in marked contrast with 2D Ising and as shown
by Fig.\ \ref{fig3}, the transition in the dLr model is from a phase
with finite correlation length to a continuous line of fixed points
[in the Renormalization Group (RG) sense], characterized by an
infinite value of the correlation length, as occurs in a KT
transition \cite{nozieres_review}.  In order to corroborate the
latter interpretation, we can try a phenomenological KT-type form
for $T^*(L)$, namely \cite{chaikin}
\begin{equation}
T^*(L)=T^{*}+\frac{a}{\left(\log L +b  \right)^2} \,,
\label{kt_fit}
\end{equation}
for constant $a$ and $b$. As seen in Fig.\ \ref{fig2}, this fit is
in very good agreement with the numerical data for large sizes. We
must caution the reader on the well-known feature of the KT
transition, that the peak of the specific heat does {\em not} occur
at the critical temperature but, rather, at a temperature preceding
$T_c$ \cite{chaikin,himbergen}. Although the size of this offset can
be model-dependent, Fig.\ \ref{fig2} indeed provides estimates, $T^*
= 1.63(1)$ on the square lattice, and $T^* = 1.85(1)$ on the
triangular lattice, that are below the corresponding $T_c$ values,
{\em and are still inside the low $T$ behavior for the spatial
correlation functions}, see Fig.\ \ref{fig3}. Thus, the inexistence
of an intermediate phase and the fact that the spatial correlations
and the specific heat change behavior at different values of $T$ can
be hardly reconciled with a single transition of the Ising class.

\begin{figure}
\begin{center}
\includegraphics[width=0.45\textwidth,clip=]{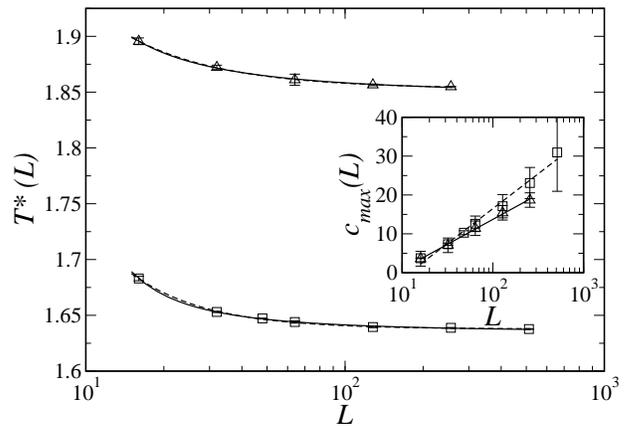}
\end{center}
\caption{Transition temperature $T^*$ as obtained from Fig.\
\ref{fig1}, as a function of $L$ for the square ($\Box$) and
triangular ($\triangle$) lattices. For each case, the dashed line is
a power-law fit $T^*(L) - T^* \sim L^{-1/\nu}$, and the solid line
is a fit to Eq.\ (\ref{kt_fit}). $L=512$ is not employed for the fit
due to low statistics. Inset: $c_{\rm max}(L)$ vs $L$ on the square
($\Box$) and triangular ($\triangle$) lattices. Lines are
logarithmic fits to the data, shown for reference. All bars
represent statistical errors and all units are arbitrary.}
\label{fig4}
\end{figure}

Our results seem to replace the KTHNY scenario for the dLr model by
a single, tensionless, KT-type phase transition. The absence of the
hexatic phase may seem surprising when contrasted with the often
accepted argument that, for increasing $T$, the surface should first
disorder in heights and, then, in slopes. However, this is a {\em
sufficient} condition for surface roughening, but it is not {\em
necessary}. For instance, in the dG model slopes are not disordered
at any temperature. It is also conceivable, as is our belief, that
heights and slopes disorder {\em at the same temperature} in the dLr
model. This remarkable result is also against the expectation that
discreteness in surface heights renormalizes the surface tension
$\sigma$, as it indeed does in the dG model \cite{nozieres_review}.
For the dLr model, generation of a non-zero $\sigma$ would imply
that the asymptotic properties of the high $T$ phase coincide with
those of the dG model \cite{note5}. In order to explore this
possibility, various analytical approaches \cite{us_anal} have been
applied to the following {\em continuum} analog of the dLr model,
introduced in the context of growth by MBE \cite{us_prl}:
\begin{equation}
\frac{\partial h}{\partial t} = - \kappa \nabla^4 h -
\frac{2\pi V}{a_{\perp}} \, \sin\left(\frac{2\pi h}{a_{\perp}}\right) +
\sqrt{2 k_B T} \, \zeta ,
\label{xMBE}
\end{equation}
where $\zeta$ is a delta-correlated Gaussian white noise, and
$a_{\perp}$, $V$, are parameters. Although a dynamical
RG study for (\ref{xMBE}) does predict the
generation of a non-zero surface tension, numerical simulations of
this Langevin equation \cite{us_prl} give results in complete
qualitative agreement with those of the dLr model presented here.
The discrepancy between the RG arguments and the numerical results
for both the discrete and continuum models might be due to
inaccuracies in the treatment of model symmetries in the RG
studies. Namely, the dLr model
can be written as a model for the {\em surface slopes} $\mathbf{m}
\equiv \nabla_d h$, i.e., ${\cal H}_{\rm dLr} = \frac{\kappa}{2}
\sum_{\mathbf{r}} [\nabla_d \cdot \mathbf{m}(\mathbf{r})]^2$, with
the implicit restriction that $\nabla_d \times \mathbf{m} \equiv
\mathbf{0}$. Thus, the dLr model has larger
symmetries than the dG model, the Hamiltonian being invariant under
arbitrary global shifts in the heights, as in the latter, but also in the slopes.
Thus the ground state degeneracy here is much larger,
minima occurring for all height configurations
for which $\nabla_d \cdot \mathbf{m} =0$. However, standard
perturbative RG analyses \cite{levin_dawson,us_anal} are oblivious
to such an added complexity in the ground state structure of the
model. Perhaps in a related fashion, the zero-vorticity
constraint for the slope field may be playing a dynamical r\^ole
in the unbinding of surface defects for $T=T_c$ in the 2D melting
transition described by the dLr model.

Summarizing, we have found that the dLr model features a single
continuous phase transition. %the hexatic phase being restricted,
%if at all, to a point in phase space.
Although sizes of our
simulations are confined to a regime in which the specific heat
still grows logarithmically with $L$, rather than saturating as in
proper KT scaling, the combined information from the spatial correlations and
the specific heat are consistent with a KT transition.
This behavior is remarkably similar to that of the continuum model
(\ref{xMBE}), including the {\em tensionless} nature of the high
$T$ phase. Progress in the analytical description of these
phenomena might improve our understanding of non-perturbative
effects in defect-mediated transitions, and of dynamical
effects of geometrical constraints (such as the curl-free
condition above) in equilibrium systems.

We thank D.\ R.\ Nelson and A.\ Sokal for discussions and A.\ S\'anchez for
his collaboration at earlier stages of this work. E.\ M.\ thanks DEAS and the
Real Colegio Complutense at Harvard for hospitality. Partial financial support
from CAM and MEC (Spain) through grants BFM2003-08532-C03-02 and FIS2004-01399
(J.\ J.\ R.-L.), BFM2003-07749-C05-01 (R.\ C.), and BFM2002-04474-C02 and
FIS2004-01001 (E.\ M.) is acknowledged. E.\ M.\ also acknowledges a Ram\'on y
Cajal contract by MEC. Part of the numerical simulations have been performed
on the BIFI PC cluster.

%\end{acknowledgments}


\begin{thebibliography}{99}
\bibitem{nota} Here, $\hat{h}(\mathbf{q})$ is the Fourier transform of
$h(\mathbf{r}) - \bar{h}$, with $\bar{h}$ the mean surface height.
\bibitem{note5} The hexatic behavior $S(q) \sim 1/(\sigma q^{2})$ is also
the asymptotics of the free Gaussian model \cite{chaikin}, and of the dG
model in the high $T$ phase, induced by surface tension.
\bibitem{nota_no_corr} A simpler minded fit to a standard scaling form (see
  Fig.\ \ref{fig2}) provides $\nu=0.71(4)$ and
$\nu=0.81(4)$ on the square and the triangular lattices, respectively. %\cite{nota_corrs}.
Besides the apparent lack of universality, these values of $\nu$, along with
our result that the specific heat exponent $\alpha\le 0$, would actually imply
violation of the hyperscaling relation \cite{chaikin} $\alpha = 2 - \nu d$,
for $d=2$.
\end{thebibliography}
\end{document}